\documentclass[preprint,12pt]{elsarticle}

\usepackage{amssymb}
\usepackage{amsmath, graphicx, mathrsfs,  amssymb, lmodern, hyperref, amscd, caption, subcaption,setspace}
\journal{Acta Astronautica}

\begin{document}

\begin{frontmatter}


 \title{\textbf{A General Quadrature Solution for Relativistic, Non-relativistic, and Weakly-Relativistic Rocket Equations}}
%\tnotetext[label1]{}
\author{Adam L. Bruce\fnref{label2}}
\ead{bruce12@purdue.edu}
\fntext[label2]{Department of Aeronautics and Astronautics, Purdue University, West Lafayette, Indiana, 47906, USA}
%%\cortext[cor1]{}
%%\address{Purdue University, West Lafayette, Indiana, 47906\fnref{label3}}

%\title{\textbf{A General Quadrature Solution for Relativistic, Non-relativistic, and Weakly-Relativistic Rocket Equations}}


%%\author{Adam L. Bruce}

%\address{Department of Aeronautics and Astronautics \\ Purdue University, West Lafayette, Indiana, 47906}

\begin{abstract}
We show the traditional rocket problem, where the ejecta velocity is assumed constant, can be reduced to an integral quadrature of which the completely non-relativistic equation of Tsiolkovsky, as well as the fully relativistic equation derived by Ackeret, are limiting cases. By expanding this quadrature in series, it is shown explicitly how relativistic corrections to the mass ratio equation as the rocket transitions from the Newtonian to the relativistic regime can be represented as products of exponential functions of the rocket velocity, ejecta velocity, and the speed of light. We find that even low order correction products approximate the traditional relativistic equation to a high accuracy in flight regimes up to $0.5c$ while retaining a clear distinction between the non-relativistic base-case and relativistic corrections. We furthermore use the results developed to consider the case where the rocket is not moving relativistically but the ejecta stream is, and where the ejecta stream is massless.   
\end{abstract}


\end{frontmatter}


\section{Introduction}
\label{intro}
A fundamental result in the theory of rocket propulsion is the relationship between the velocity increment a rocket gains and the amount of reaction mass it expels to gain this increment. The canonical form of this equation, which is credited to Tsiolkovsky, is derived under the assumption that the rocket is moving slowly compared to the speed of light throughout its flight (\cite{sutton}), and thus its momentum is given by the Newtonian formula $\mathbf{p} = m_R\mathbf{v}$. Another relationship between rocket velocity and expelled reaction mass governs the flight of rockets operating in the relativistic regime. The relativistic rocket equation was first derived by Ackeret (\cite{Ackeret}) in 1946, and subsequently expounded upon by Krause (\cite{krause}), Bade (\cite{bade}), and Forward (\cite{forward},\cite{spad}), the last of whom presented a ``transparent derivation'' of the relativistic rocket equation based on simple equations in special relativity. Other research has been done in relativistic rocketry, such as simultaneity calculations based on the propagation of light signals to and from a constantly accelerating rocket (\cite{greenwood}), calculations of the maximum attainable velocity for a relativistic rocket (\cite{maxvel}) and how an $I_{sp}$ like parameter could be constructed for a rocket using exotic propellants whose products consist of both massive and massless particles (\cite{noteorig},\cite{note}). 

One issue which has not yet been investigated is how a rocket will perform in a weakly relativistic regime, and how quasi post-Newtonian corrections can be derived for the Tsiolkovsky equation. In fact, while it has been noted by Forward (\cite{forward}) that the Ackeret rocket equation converges to the Tsiolkovsky equation if evaluated numerically, the explicit proof of this fact does not appear to have been previously published. We shall show that starting from the same simple equations as Forward it is possible to reduce the rocket equation to a general quadrature, without making any assumptions about its flight regime. Moreover, this quadrature is shown to reproduce the equations of both Tsiolkovsky and Ackeret when evaluated in the correct limits. Finally it is shown that by expanding this quadrature in series it is possible to obtain a product of quasi post-Newtonian corrections which converges to the Ackeret equation in the limit as $n\rightarrow\infty$. Section 2 presents the derivation of the main result of this work, Equation (\ref{eq:main}), while Section 3 shows how the Tsiolkovsky and Ackeret equations as well as the post-Newtonian terms can be all derived from this result. Section 4 uses the results we develop to analyze two seemingly novel scenarios, which are found to yield results that are surprisingly consistent with a naive application of the Ackeret's relativistic equation.

\section{The Rocket Problem and its Reduction to Quadrature}
In this section we discuss some basic relativistic principles necessary for the derivation and show how the quadrature is derived. 

\subsection{Assumptions and Definitions}
Just as energy and momentum are conserved separately in Newtonian mechanics, energy-momentum is conserved as a 4-vector in special relativity. We define the energy-momentum of a system in (1+1)-dimensions as%\footnote[\]{We use the index convention of taking all momenta and velocities as canonically covariant, in principle leaving all possible position variables as canonically contravariant. As we shall find the distinction between these two types of indices is immaterial as no contravariant quantities appear in the derivation.} 
\begin{equation}
p_{\mu}:(p_0,p_1)=(E/c,p).
\end{equation}
Similarly, we may define a (1+1)-dimensional 4-velocity which is given by
\begin{equation}
u_{\mu}:(u_0,u_1)=(\gamma c,\gamma v),
\end{equation}
where $\gamma$ is the Lorentz factor of the frame which is defined to be
\begin{equation}
\gamma \equiv \frac{1}{\sqrt{1-v^2/c^2}}.
\end{equation}
One may note that in this form the familiar Newtonian relations still hold, viz.
\begin{equation}
p_{\mu} = mu_{\mu},
\end{equation}
which the reader may verify. 

In order to perform any calculation in relativistic mechanics we must first specify a frame for the observer, which will be considered to be at rest. The most convenient frame for a rocket calculation will be the Center-of-Momentum (COM) frame, which is defined as the unique frame for which 
\begin{equation}
\sum_{j} (p_{\mu})_j = 0.
\end{equation}
It can be shown in Newtonian mechanics that the center of mass of any closed system possesses a constant velocity. A similar conservation principle holds in special relativity and the COM frame may be thought of as the unique inertial frame which is comoving with the center of mass and whose origin coincides with the position of the center of mass.

\subsection{The Rocket Problem}
In the rocket problem, a body which is initially moving with velocity $v_o$ (which is zero in the COM frame) ejects an infinitesimal amount of mass $dm$ which propels the body forward in accordance with the conservation of momentum. In the relativistic picture this is equivalent to the conservation of energy-momentum as has been defined in Section 2.1. In order to form the energy-momentum balance we consider the rocket at two infinitesimally separated instants: 
\begin{figure}
\begin{center}
\includegraphics[scale=0.4]{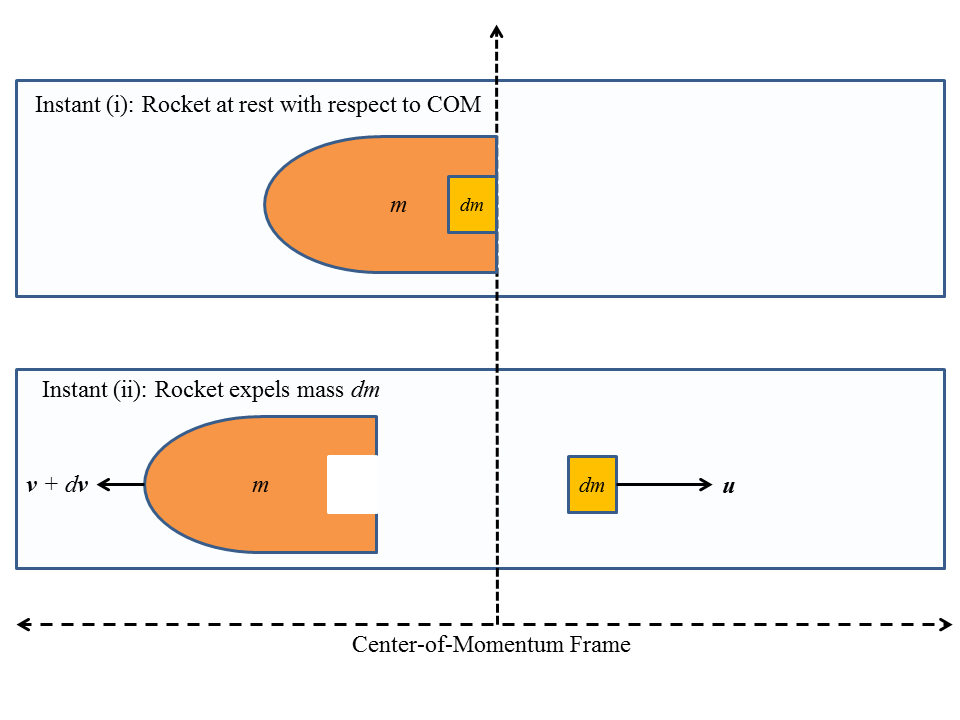}
\end{center}
\caption{Schematic of the energy-momentum balence.}
\end{figure}
\vspace{5mm}

\noindent(i): The rocket possesses zero velocity with respect to the COM frame and has rest mass $m+dm$. We shall consider $dm$ to be the rest mass of the propellants so that after they are expelled the rocket possesses a rest mass $m$.

\vspace{5mm}

\noindent(ii): The rocket expels the mass $dm$ and gains a velocity increment $d\mathbf{v}$. This separates the center of mass of the rocket from the center of momentum of the system and thus gives the rocket a nontrivial velocity with respect to the COM frame. 

\vspace{5mm}
\noindent The process is further illustrated in Figure 1. As stated before, we may write energy-momentum conservation in the COM frame as 
\begin{equation}\label{eq:syscon}
d\left(\sum_{j}(p_{\mu})_j\right) = \sum_{j}(dp_{\mu})_j = 0.
\end{equation}
For the rocket system the energy-momentum consists of two components, the rocket component $p_\mu^R$ and the ejecta component $p_{\mu}^e$. Substituting this into Equation (\ref{eq:syscon}) we find
\begin{equation}
dp_\mu^R = -dp_{\mu}^e.
\end{equation}
Because the ejecta is presumed to be at rest to begin with, the differential momentum for the ejecta is equal to the final momentum, which is given by $u_{\mu}dm$, so the general energy momentum balance is given by
\begin{equation}\label{eq:rbal}
d(E_R/c,\mathbf{p}_R) = -dm(\gamma_e c,\gamma_e \mathbf{u}),
\end{equation}
where $\mathbf{u}$ is the ejecta velocity in the center-of-momentum frame. The rocket's total energy is given by $E_R = \gamma_R m_R c^2$, while its total momentum is given by $\mathbf{p}_R = \gamma_R m_R \mathbf{v}$. Substituting these into Equation (\ref{eq:rbal}) along with the signs for the 1-dimensional vectors and separating the equations gives the system
\begin{equation}\label{eq:sys}
\begin{aligned}
d(\gamma_R m_R) &= -\gamma_e dm,\\
d(\gamma_R m_R v) &= \gamma_e u dm.\\
\end{aligned}
\end{equation}
The first of these equations represents the conservation of energy while the second represents the conservation of momentum. Forward (\cite{forward}) at this point begins to evaluate the derivatives on the left hand side of the equation and upon substituting the fully relativistic form of $u$ into the momentum equation arrives ultimately at Ackeret's equation, unfortunately only by proceeding through a forest of differentiation and algebraic reduction. Suppose instead we substitute the first equation into the second to give
\begin{equation}
d(\gamma_R m_R v) = -ud(\gamma_R m_R).
\end{equation}
The derivative on the left hand side of this equation is 
\begin{equation}
d(\gamma_R m_R v) = vd(\gamma_R m_R)+\gamma_R m_Rdv,
\end{equation} 
so defining
\begin{equation}
\mu \equiv \gamma_R m_R,
\end{equation}
we may write this equation as
\begin{equation}\label{eq:maindiff}
(v+u)d\mu+\mu dv = 0.
\end{equation}
Equation (\ref{eq:maindiff}) is a single, separable differential equation which describes any kind of rocket motion regardless of the velocity regime. Without specifying the expression to be used for $u$, we simply leave $u=u(v)$ and perform the integration to find
\begin{equation}\label{eq:middlemain}
\ln\frac{\mu_0}{\mu} = \int_{0}^{v}\frac{dv'}{v'+u(v')}.
\end{equation}
Since $v(t=0)=0$, we have $\gamma(t=0) = 1$ and thus $\mu_0 = m_0$. Therefore we may write equation (\ref{eq:middlemain}), restoring the Lorentz factors and defining the typical mass ratio ($\text{MR} \equiv m_\text{initial}/m_\text{final}$), as
\begin{equation}\label{eq:main}
\text{MR} = \gamma e^{\Omega(v)} \hspace{3mm} \text{where} \hspace{3mm} \Omega(v) \equiv \int_{0}^{v}\frac{dv'}{v'+u(v')}.
\end{equation}
Equation (\ref{eq:main}) forms the main result of the section and also the titular result of this work. In the next section we shall show how both the Tsiolkovsky equation and Ackeret's relativistic equation emerge in the limit where $u$ is treated Newtonianly and relativistically, respectively, as well as show that a series expansion of the integrand can be used to obtain relativistic corrections to a rocket which is flying in an intermediate regime.

\section{Evaluation of the Quadrature Solution in Various Flight Regimes}
The evaluation of Equation (\ref{eq:main}) is accomplished by specifying a functional form for $u(v)$ and performing the integration. We shall consider three such cases in this section corresponding in order to the Newtonian regime, relativistic regime, and the transition between the two. 
\subsection{The Newtonian and Relativistic Regimes}
If $v$ is far less than $c$, the relative velocity $u$ is obtained through Newtonian addition of velocities,
\begin{equation}
u = v_e-v,
\end{equation}
and the factor $v/c$ is so small that $\gamma = 1$ to any reasonable number of significant figures. Thus we have
\begin{equation}
\Omega(v) = \int_{0}^{v}\frac{dv'}{v'+(v_e-v')}=\frac{v}{v_e},
\end{equation} 
so 
\begin{equation}
\text{MR} = \gamma e^{\Omega(v)} = e^{v/v_e},
\end{equation}
which is Tsiolkovsky's equation. 

Conversely, if $v$ is close to $c$ %, perhaps $0.5c$ or above\footnote{We include the possiblity of large relativistic velocities merely for the sake of argument, as it has been shown by [cite people] that the limit to the velocity any realistic rocket could achieve which is below the regime of $0.9c$.}, 
we must use the fully relativistic velocity addition formula, which is given by
\begin{equation}
u = \frac{v_e-v}{1-\beta_e\beta},  \hspace{3mm} \text{where} \hspace{3mm} \beta_i \equiv v_i/c.
\end{equation}
Substituting this into Equation (\ref{eq:main}) and simplifying the expression gives 
\begin{equation}
\Omega(v) = \frac{1}{v_e}\int_{0}^{\beta}\frac{1-\beta_e\beta'}{1-\beta'^2}(cd\beta')=\frac{1}{\beta_e}\int_{0}^{\beta}\frac{d\beta'}{1-\beta'^2}-\int_{0}^{\beta}\frac{\beta'}{1-\beta'^2}d\beta'.
\end{equation}
The rightmost integration is simple to perform and gives
\begin{equation}
\int_{0}^{\beta}\frac{\beta'}{1-\beta'^2}d\beta' = -\frac{1}{2}\ln(1-\beta^2),
\end{equation}
serving to effectively cancel out the factor of $\gamma$ in front of the exponential. The second integral is more difficult, but is found to be
\begin{equation}\label{eq:int}
\int_{0}^{\beta}\frac{d\beta'}{1-\beta'^2} = \frac{1}{2}\ln\frac{1+\beta}{1-\beta},
\end{equation}
which, when exponentiated, gives the familiar equation of Ackeret,
\begin{equation}\label{eq:ack}
\text{MR} = \left(\frac{1+\beta}{1-\beta}\right)^{\frac{1}{2\beta_e}}.
\end{equation}
This shows Equation (\ref{eq:main}) gives the correct results for both the relativistic and Newtonian limits.

\subsection{The Transition to Relativistic Flight: A Series Solution}
Another result can be obtained from Equation (\ref{eq:main}) if we expand the integrand of Equation (\ref{eq:int}) in series, viz.
\begin{equation}
\frac{1}{1-\beta^2} = \sum_{\nu=0}^{\infty}\beta^{2\nu}.
\end{equation}
Integration of this series is trivial, producing 
\begin{equation}
 \int_{0}^{\beta}\frac{1}{1-\beta'^2}d\beta' = \int_{0}^{\beta} d\beta'\sum_{\nu=0}^{\infty}\beta'^{2\nu}=\sum_{\nu=0}^{\infty}\frac{\beta^{2\nu+1}}{2\nu+1},
\end{equation}
which gives the expression for the relativistic mass ratio in terms of a product,
\begin{equation}\label{eq:infprod}
\text{MR} = \prod_{\nu=0}^{\infty}\exp\left(\frac{1}{\beta_e}\frac{\beta^{2\nu+1}}{2\nu+1}\right) = e^{v/v_e}\prod_{\nu=1}^{\infty}\exp\left(\frac{1}{\beta_e}\frac{\beta^{2\nu+1}}{2\nu+1}\right).
\end{equation}
If the product is taken to infinity, the final result will be the same as Ackeret's equation. Indeed, it can be shown that if the leftmost product is reconstructed as the exponentiation of a series, this series converges exactly to the result of Equation (\ref{eq:int}), a trivial result given the simple steps taken to derive this form of the equation.

\begin{figure}[t]
\begin{center}
\includegraphics[scale=0.45]{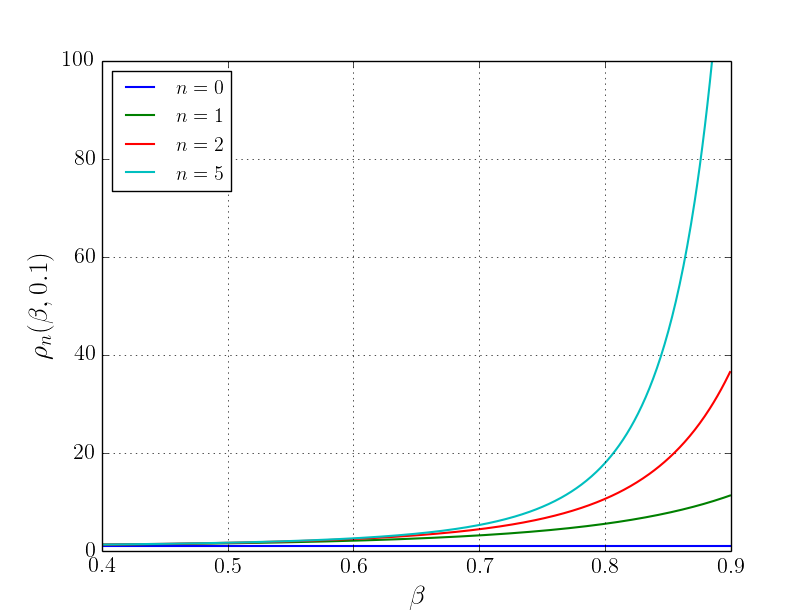}
\caption{A plot of $\rho_n(\beta ;\beta_e=0.1)$ for $n = 0$ (non-relativistic)$,1,2$ and $5$.}
\end{center}
\end{figure}

Let us consider the case where $\beta$ is large enough so that at least the first term of the rightmost product would not be unity, but not so large that an infinite number of terms need be accounted for to successfully approximate the mass ratio. In this case it is appropriate to truncate the product in Equation (\ref{eq:infprod}) to a few low-order terms. Such truncations of this product show how the mass ratio varies as the rocket begins to enter the relativistic regime. Let us consider the family of rocket equations
\begin{equation}
\text{MR}_n = e^{v/v_e}\rho_n(\beta ;\beta_e),
\end{equation}
where
\begin{equation}
\rho_n(\beta ;\beta_e) \equiv\prod_{\nu=1}^{n}\exp\left(\frac{1}{\beta_e}\frac{\beta^{2\nu+1}}{2\nu+1}\right) \hspace{3mm} \text{and} \hspace{3mm} \rho_o = 1.
\end{equation}

For the convenience of the reader, the first few $\rho_n(\beta ;\beta_e)$ are given in Table 1 in terms of $v$, $v_e$, and $c$ and a plot of $\rho_n(\beta ;\beta_e=0.1)$ for $n = 0,1,2$ and $5$ is shown in Figure 2.
 
\begin{table}[b]\label{tab:terms}
\begin{center}
\begin{tabular}{||c|c||}
\hline
\hline
$n$ & $\rho_n(v;v_e)$\\
\hline
0 & 1\\

1 & $e^{v^{3}/3c^2v_e}$\\

2 & $e^{v^{3}/3c^2v_e}e^{v^{5}/5c^4v_e}$\\

3 & $e^{v^{3}/3c^2v_e}e^{v^{5}/5c^4v_e}e^{v^{7}/7c^6v_e}$\\

4 & $e^{v^{3}/3c^2v_e}e^{v^{5}/5c^4v_e}e^{v^{7}/7c^6v_e}e^{v^{9}/9c^8v_e}$\\

5 & $e^{v^{3}/3c^2v_e}e^{v^{5}/5c^4v_e}e^{v^{7}/7c^6v_e}e^{v^{9}/9c^8v_e}e^{v^{11}/11c^{10}v_e}$ \\
\hline

\end{tabular}
\end{center}
\caption{Relativistic corrections for the Tsiolkovsky equation up to 5th order.}
\end{table}
The set of equations denoted by $\text{MR}_n$ include the Tsiolkovsky equation ($\text{MR}_0$) and Ackeret's equation ($\text{MR}_\infty$). As the rocket transitions from the completely Newtonian regime to a weakly relativistic regime, the mass ratio stops being represented by $\text{MR}_0$ and begins being represented by $\text{MR}_1$, given by
\begin{equation}
\text{MR}_1 = e^{v/v_e}\rho_1(\beta ;\beta_e) = e^{v/v_e}e^{v^{3}/3c^2v_e}.
\end{equation}
The effect of this term is to require more propellant than would be required by the Tsiolkovsky equation alone to achieve the same final velocity were relativistic effects neglected. As the rocket moves still faster the second term in the product becomes significant and the mass ratio equation becomes 
\begin{equation}
\text{MR}_2 = e^{v/v_e}e^{v^{3}/3c^2v_e}e^{v^{5}/5c^4v_e},
\end{equation} 
\begin{figure}[h]
\centering
\begin{subfigure}{0.59\textwidth}
\includegraphics[width=\textwidth]{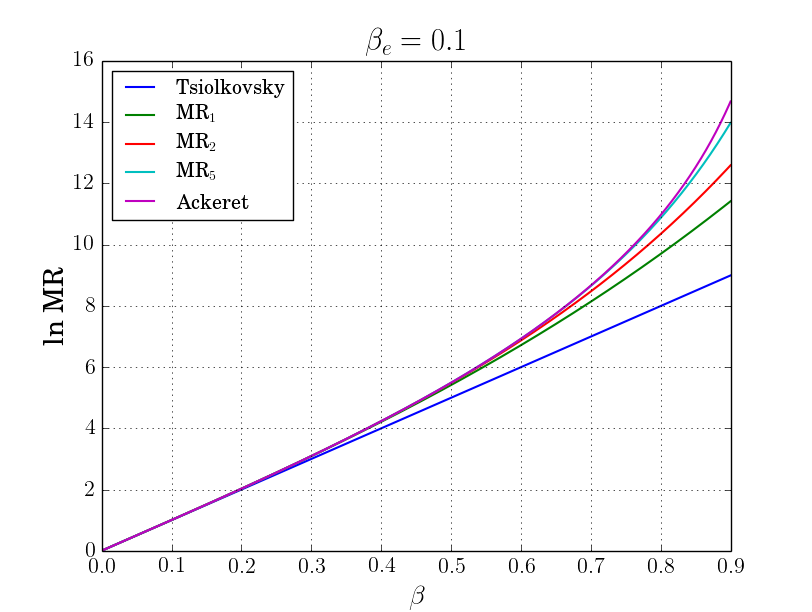}
\end{subfigure}

\begin{subfigure}{0.59\textwidth}
\includegraphics[width=\textwidth]{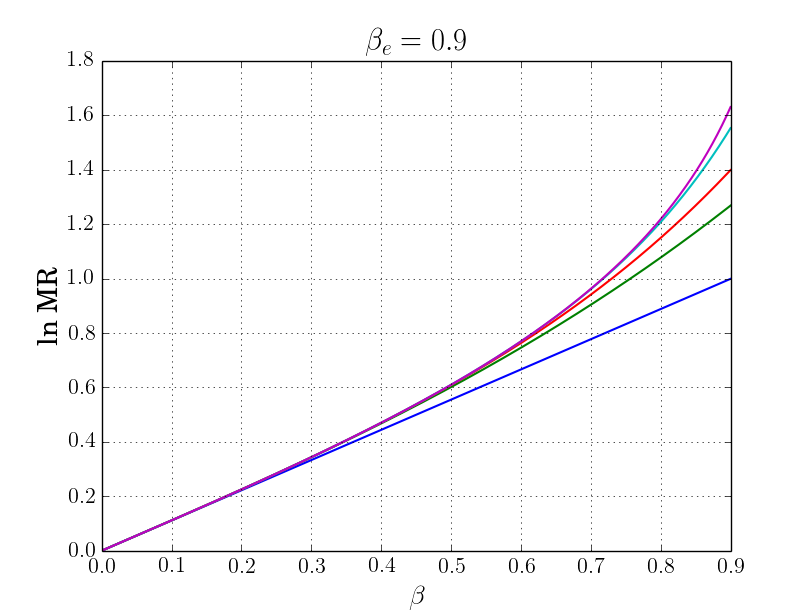}
\end{subfigure}

%\subfigure{
%\rule{4cm}{3cm}
%\includegraphics[width=\textwidth]{mrcompbeta0p1.png}
%}
%
%\subfigure{
%\rule{4cm}{3cm}
%\includegraphics[width=\textwidth]{mrcompbeta0p5.png}
%}
%
%\subfigure{
%\rule{4cm}{3cm}
%\includegraphics[width=\textwidth]{mrcompbeta0p9.png}
%}
\caption{Mass Ratios versus $\beta$ for various MR$_n$ equations and $\beta_e$. The actual value of $\beta_e$ does not change the relative morphology of the curves.}
\end{figure}
and so on until so many terms are required that it is simpler to use Ackeret's equation. An obvious course of action from here is to determine the velocity range over which these low order truncations of the infinite product accurately predict the mass ratio. The difference between using $\text{MR}_1$ and Equation (\ref{eq:ack}) for $\beta_e = 0.5$ and $\beta = 0.4$ is about $0.5\%$, and as little as $0.3\%$ for $\beta_e = 0.999$ and $\beta = 0.4$. At much larger values of $\beta$, the first few terms of this product are no longer sufficient, and it would be much easier to use the full relativistic equation. Figure 3 shows plots of $\ln\text{MR}$ versus $\beta$ given a fixed $v_e$ for various $n$ as well as the Tsiolkovsky equation with Ackeret's equation as a limiting case. Note that the relative morphology of the curves, and thus the extent to which $\text{MR}_n$ for a given $n$ approximates the Ackeret equation, is invariant with respect to $\beta_e$.

\section{Related Questions}
In the previous sections we showed the rocket equation can be reduced to a unique quadrature which is not dependent on the flight regime which is considered, and have used this fact to show explicitly the transition between Newtonian and relativistic flight regimes. In this section we discuss further questions which can be answered by the results we have developed, in particular the case where $v_e$ is relativistic but $v$ is not and the case where the ejecta stream is massless but still has momentum.

\subsection{Relativistic Ejecta Stream on a Newtonian Rocket}
If $\beta_e$ is considerable but $\beta \ll 1$ we may write the equations of (\ref{eq:sys}) as 

\begin{equation}
\begin{aligned}
dm_R &= -\gamma_e dm,\\
d(m_R v) &= \gamma_e u dm.\\
\end{aligned}
\end{equation}
This gives the differential equation
\begin{equation}
udm_R + d(m_R v) = 0,
\end{equation}
which is indistinguisable from the classical case. In this case the gain in rocket momentum from the relativistic divergence of the ejecta momentum is exactly counteracted by the divergence in relativistic kinetic energy which the ejecta stream must gain before exiting the vehicle. This also explains the invariance of the relative morphology between the Ackeret equation and various $\text{MR}_{n}$ with respect to $\beta_e$ which was shown in the last section. 

\subsection{Rocket Equation for A Massless Ejecta Stream}
Although the practicality of using a massless ejecta stream is disputed (\cite{note}), such cases have been considered in the literature before (\cite{noteorig},\cite{forantipart},\cite{photon}). A rocket equation governing this case may be derived from Equations (\ref{eq:sys}) and Equation (\ref{eq:main}) alike. For the analysis of this case we shall consider massive reactants which are reacted in such a way as to create massless products that form the ejecta stream. This in particular ensures that the interpretation of the mass ratio remains obvious as the ratio of the total mass before and after the burn. 

In relativity the total energy of an object (rest-energy and kinetic energy) is given by
\begin{equation}
E = \sqrt{(\mathbf{p}\cdot\mathbf{p})c^2+(mc^2)^2}.
\end{equation}
Basic quantum mechanical principles allow a second relationship between particle momentum and quantum frequency,
\begin{equation}\label{eq:qeng}
E = \hbar \omega,
\end{equation}
so for a massless particle of quantum frequency $\omega$, 
\begin{equation}\label{eq:qmom}
p = \frac{\hbar\omega}{c}.
\end{equation}
The existence of nontrivial energy and momentum relations implies it is possible in theory to create thrust using massless propellants alone, and thus there should be a well-defined rocket equation governing such propellants. Furthermore it is known that massless particles (e.g. photons) travel at the speed of light regardless of the frame in which they are observed, and thus for the case of such particles $u(v) = v_e = c$. Equation (\ref{eq:main}) then gives
\begin{equation}
\Omega(v) = \int_{0}^{v}\frac{dv'}{v'+c}= \ln\left(1+\beta\right),
\end{equation} 
which gives the rocket equation
\begin{equation}
\text{MR} = \sqrt{\frac{1+\beta}{1-\beta}}.
\end{equation}
This same result may also be obtained by using the energy and momentum relations of Equations (\ref{eq:qeng}) and (\ref{eq:qmom}) and writing a system of equations similar to the equations (\ref{eq:sys}) for a large number of massless particles. It is however more straightforward to perform the calculation using Equation (\ref{eq:main}). Furthermore this exactly agrees with Ackeret's equation in the case where $\beta_e = 1$, an interesting result considering Ackeret's equation is derived under the assumption of the ejecta stream having a finite, nonzero mass.

\section{Conclusions}
The dynamics involved in rocket motion are simple, and rely on basic conservation principles which are obeyed in every possible flight regime, including those which are not practically available due to technological limitations. This idea itself suggests that there should be a generic form the rocket equation, whose Newtonian and relativistic limits reproduce known results. We have shown that such a generic equation exists and can be simply derived from basic considerations. We have also used this generic equation to derive a series expansion of quasi post-Newtonian corrections which show the transition from non-relativistic to relativistic flight, and unexpectedly found that the traditional relativistic rocket equation is well approximated up to $0.5c$ by even the first-order relativistic correction to the Tsiolkovsky equation. Furthermore, we have found the difference in morphology between the traditional relativistic equation, Tsiolkovsky equation, and Tsiolkovsky equation with some number of relativistic expansion terms included is invariant with respect to the regime of the ejecta velocity. The most striking case of this is the finding that a rocket traveling non-relativistically but utilizing a relativistic ejecta stream will not find any relativistic discrepancy, even in theory, between its true mass ratio and that predicted by the Tsiolkovsky equation. We have also found that the traditional relativistic equation will still accurately govern the mass ratio under the assumption that the ejecta stream is massless. 

It is possible that still further scenarios can be easily analyzed by formulating design-specific expressions for $u(v)$ and using the generic rocket equation we have derived to understand the mass ratio equation which results from a given design. Examples of such scenarios might be hypothetical rockets using multiple ejecta streams of differing velocities, and perhaps operating within different velocity regimes. 

\section*{Acknowledgements}
I am indebted to Professors J. Longuski and E. Fischbach, of the Department of Aeronautics and Astronautics and Department of Physics at Purdue University respectively, for hearing and offering guidance on the decision to submit this work for publication. I am also deeply grateful to my dear friend and colleague J. Smolinsky, of the Department of Physics at the University of California, Irvine, for his insightful discussion of and contributions to the relativistic formulation of the rocket problem here presented.






\end{document}